\documentclass[sigplan,10pt,nonacm]{acmart}

\renewcommand\footnotetextcopyrightpermission[1]{}
\usepackage{booktabs}
\usepackage{subcaption}
\usepackage{xcolor}
\usepackage{todonotes}
\usepackage{soul}
\usepackage{graphicx}
\usepackage{tikz}
\usetikzlibrary{positioning,decorations.pathreplacing,fit,backgrounds}
\usepackage{placeins}
\usepackage{colortbl}
\usepackage{enumitem}
\usepackage{algorithm}
\usepackage{algpseudocode}
\definecolor{cblue}{RGB}{55,131,187}
\definecolor{cred}{RGB}{211,63,73}
\definecolor{cgreen}{RGB}{56,158,92}
\definecolor{corange}{RGB}{230,159,0}

\newif\ifsubmission
\submissionfalse    

\ifsubmission
\newcommand{\mcnote}[1]{}
\newcommand{\jhnote}[1]{}
\newcommand{\tlnote}[1]{}
\else
\newcommand{\mcnote}[1]{\todo[color=violet!40,inline]{\hl{\textbf{MC:} } #1}}
\newcommand{\jhnote}[1]{\todo[color=cyan!10!white,inline]{\hl{\textbf{JH:} }\small #1}}
\newcommand{\tlnote}[1]{\todo[color=green!15!white,inline]{\hl{\textbf{TL:} }\small #1}}
\fi

\title[Why Smaller Is Slower? Dimensional Misalignment in Compressed LLMs]{Why Smaller Is Slower?\\ Dimensional Misalignment in Compressed LLMs}

\author{Jihao Xin}
\affiliation{
  \institution{KAUST}
  \country{}
}

\author{Tian Lyu}
\affiliation{
  \institution{KAUST}
  \country{}
}

\author{Qilong Pan}
\affiliation{
  \institution{HUMAIN}
  \country{}
}

\author{Kesen Wang}
\affiliation{
  \institution{HUMAIN}
  \country{}
}

\author{Marco Canini}
\affiliation{
  \institution{KAUST}
  \country{}
}

\begin{abstract}
Post-training compression reduces LLM parameter counts but often produces irregular tensor dimensions that degrade GPU performance---a phenomenon we call \emph{dimensional misalignment}.
We present a full-stack analysis tracing root causes at three levels: framework, library, and hardware.
The key insight is that model inference becomes slower because the resulting dimensions are unfriendly with the GPU execution stack.
For example, compressing Llama-3-8B with activation-aware singular
value decomposition (ASVD) has 15\% fewer parameters yet runs no faster than the uncompressed baseline, because 95\% of its dimensions are misaligned.

We propose \textbf{GAC} (GPU-Aligned Compression), a new compression paradigm that wraps any dimension-reducing compressor and re-selects hardware-aligned dimensions via multi-choice knapsack optimization under the same parameter budget.
We evaluate GAC on Llama-3-8B with ASVD and LLM-Pruner, achieving 100\% alignment and recovering up to 1.5$\times$ speedup while preserving model quality.
\end{abstract}

\begin{document}

\maketitle

\section{Introduction}
\label{sec:intro}

Large Language Models (LLMs) deliver strong capabilities, but their scale hinders deployment.
Post-training compression reduces model size and can be categorized into three families: \emph{quantization}~\citep{gptq,awq}, \emph{sparsification}~\citep{sparsegpt,wanda}, and \emph{dimension reduction}~\citep{asvd,palu,pyramidkv}.
Quantization and sparsification preserve tensor shapes and are therefore compatible with existing GPU kernels.
Pretrained models come with carefully designed dimensions that are GPU-friendly, such as 8-aligned dimensions for FlashAttention~\citep{flashattention}.
Dimension reduction, however, alters the tensor shapes, often producing \emph{irregular} dimensions (e.g., dimension reduced from 128 to 107), which conflict with GPU execution primitives---Tensor Core tiles, cuBLAS kernels, and framework backend selection---causing a counterintuitive outcome: \emph{models with \textbf{fewer} parameters run \textbf{slower} than uncompressed ones}.
We term this phenomenon \textbf{dimensional misalignment}.

\begin{figure}[t]
\centering
\resizebox{0.85\columnwidth}{!}{%
\begin{tikzpicture}[
    dim/.style={<->, line width=0.7pt, gray!80!black},
    dlabel/.style={font=\sffamily\bfseries, fill=white, inner sep=1pt}]
  \fill[cblue!8] (0,0) rectangle (1.6,2.6);
  \draw[line width=0.8pt] (0,0) rectangle (1.6,2.6);
  \node at (0.8,1.3) {\large $\mathbf{X}$};
  \draw[dim] (-0.4,0) -- (-0.4,2.6) node[midway, dlabel] {$M$};
  \draw[dim] (0,3.0) -- (1.6,3.0) node[midway, dlabel] {$K$};
  \node at (2.4,1.8) {\LARGE $\cdot$};
  \fill[cgreen!8] (3.2,1.0) rectangle (5.8,2.6);
  \draw[line width=0.8pt] (3.2,1.0) rectangle (5.8,2.6);
  \node at (4.5,1.8) {\large $\mathbf{W}$};
  \draw[dim] (2.9,1.0) -- (2.9,2.6) node[midway, dlabel] {$K$};
  \draw[dim] (3.2,3.0) -- (5.8,3.0) node[midway, dlabel] {$N$};
  \node at (6.6,1.3) {\LARGE $=$};
  \fill[corange!8] (7.4,0) rectangle (10.0,2.6);
  \draw[line width=0.8pt] (7.4,0) rectangle (10.0,2.6);
  \node at (8.7,1.3) {\large $\mathbf{Y}$};
  \draw[dim] (7.1,0) -- (7.1,2.6) node[midway, dlabel] {$M$};
  \draw[dim] (7.4,3.0) -- (10.0,3.0) node[midway, dlabel] {$N$};
\end{tikzpicture}%
}
\caption{GEMM: $Y{=}X{\cdot}W$ with $X{\in}\mathbb{R}^{M \times K}$,
$W{\in}\mathbb{R}^{K \times N}$.}
\label{fig:GEMM_dims}
\end{figure}
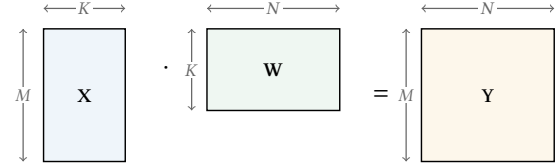

To see why, consider one of the core operators in LLMs, GEMM (General Matrix Multiply) in the expression: $Y{=}X{\cdot}W$ (Figure~\ref{fig:GEMM_dims}), with sequence dimension~$M$, inner dimension~$K$, and output dimension~$N$.
Three compression methods each target a different axis:
\textbf{(i)}~low-rank factorization replaces $W$ with two factors $A{\cdot}B$, introducing a reduced rank $r$ as the inner dimension of two smaller GEMMs~\citep{asvd,palu,svdllm2024};
\textbf{(ii)}~structured pruning removes neurons or attention heads, shrinking~$N$~\citep{llmpruner};
\textbf{(iii)}~token/KV eviction drops sequence entries, reducing~$M$~\citep{h2o,pyramidkv}.
All three optimize \emph{size vs.\ accuracy} without checking whether the resulting dimensions satisfy GPU alignment constraints (e.g., whether $d \bmod 8 = 0$).
The problem is pervasive: ASVD~\citep{asvd} produces 95\% misaligned dimensions, and even LLM-Pruner~\citep{llmpruner}---whose pruning granularity is relatively coarse---still leaves 17\% of weights misaligned (Table~\ref{tab:main_results}).
Prior hardware-aware methods~\citep{halp2021,haloc2023} treat latency as a black-box, tying solutions on specific architectures such as CNNs without isolating root causes or providing parameter-budget guarantees.

We propose \textbf{GAC} (GPU-Aligned Compression), a compression \emph{paradigm} that imposes hardware alignment constraints on any dimension-reducing compressor.
GAC first identifies \emph{why} certain dimensions are slow via a full-stack analysis (\S\ref{sec:analysis}); next, it constrains a multi-choice knapsack optimizer to re-select aligned dimensions under the same parameter budget.
GAC is a compressor-agnostic framework that restores speed without changing the upstream compressor's design.

\paragraph{Contributions.}
\begin{itemize}
    \item We identify the \emph{dimensional misalignment} problem in compressed LLMs and demonstrate its prevalence across compressors (\S\ref{sec:misalignment}).
    \item We conduct a systematic full-stack analysis to trace the root causes and categories them into three layers: framework, library, and hardware(\S\ref{sec:analysis}).
    \item We formalize the dimension selection as a constrained optimization and provide a dynamic programming solver to guarantee the alignment (\S\ref{sec:gac}).
    \item Preliminary evaluation on Llama-3-8B~\citep{llama3} achieves up to 1.5$\times$ speedup without sacrificing accuracy (\S\ref{sec:eval}).
\end{itemize}

\section{Motivation: Dimensional Misalignment}
\label{sec:misalignment}
\begin{figure*}[t]
  \centering
  \includegraphics[width=\textwidth]{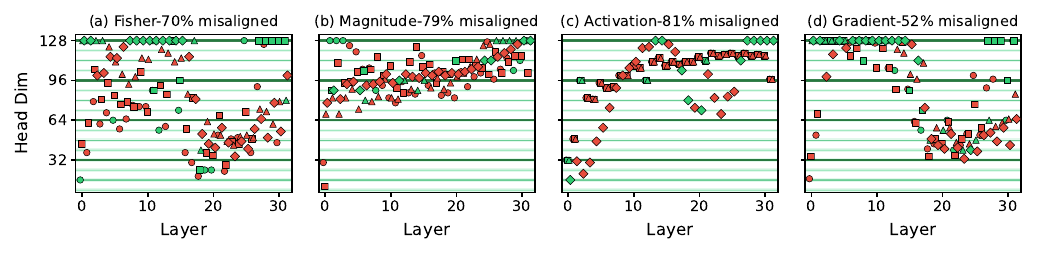}
  \caption{Llama-3-8B at $\rho=20\%$. Shape: $\circ{=}$Q Head, $\square{=}$K Head, $\triangle{=}$V Head; color: \textcolor{cgreen}{green}=8-aligned, \textcolor{cred}{red}=misaligned.}
  \label{fig:dim_scatter}
  \end{figure*}

Existing LLM compression aims to preserve accuracy under a given compression ratio $\rho$.
Formally, given the set of pretrained parameters $\mathcal{W}$, we seek a compressed set $\mathcal{W}'$ that minimizes expected loss subject to the size constraint:
\begin{equation}
\min_{\mathcal{W}'} \; \mathbb{E}_{(x,y)\sim\mathcal{D}}[\mathcal{L}(\mathcal{W}'; x, y)] \quad \text{s.t.} \quad 1-|\mathcal{W}'|/|\mathcal{W}| \geq \rho.
\label{eq:compression_target}
\end{equation}
Here $\mathcal{D}$ is the data distribution, $\mathcal{L}$ is the loss, $|\mathcal{W}'|$ and $|\mathcal{W}|$ denote total parameter counts. We denote $B = (1-\rho)|\mathcal{W}|$ as the parameter budget.

Different parameters have different compression sensitivities, e.g., early and late layers are often more critical than middle layers---so budget cannot be allocated uniformly.
Existing methods proceed in two steps.
First, for each parameter $W_i$, compute an \textbf{importance score} $s_i$ using a proxy (Table~\ref{tab:importance_scores}); a higher $s_i$ reflects higher sensitivity, so we should retain more parameters in $W_i$.
Second, allocate the budget $B$ by assigning each $W_i$ a dimension $d_i$ so that more important $W_i$ receive larger $d_i$:
\begin{equation}
\{d_i\} = \arg\max_{\{d_i\}} \sum_i s_i \cdot |W_i| \quad \text{s.t.} \quad |\mathcal{W}'| \leq B, \quad d_i \geq 0.
\label{eq:budget_allocation}
\end{equation}
Here $d_i$ is the compressed dimension (e.g., inner dimension or output width, varies by compression method), $|W_i|$ is the parameter count given dimension $d_i$, $|\mathcal{W}'| = \sum_i |W_i|$ is the total parameter count of the compressed model, and $B$ is the parameter budget.
Because $s_i$ and the optimum $\{d_i\}$ are continuous, the resulting dimensions are typically \emph{irregular} (e.g., 107, 108, 109) and often violate GPU alignment requirements (e.g.\ $d_i \bmod 8 = 0$),
 triggering backend fallbacks and kernel switches that erase the expected speedup from fewer parameters and FLOPs.

\begin{figure}[t]
\centering
\includegraphics[width=\columnwidth]{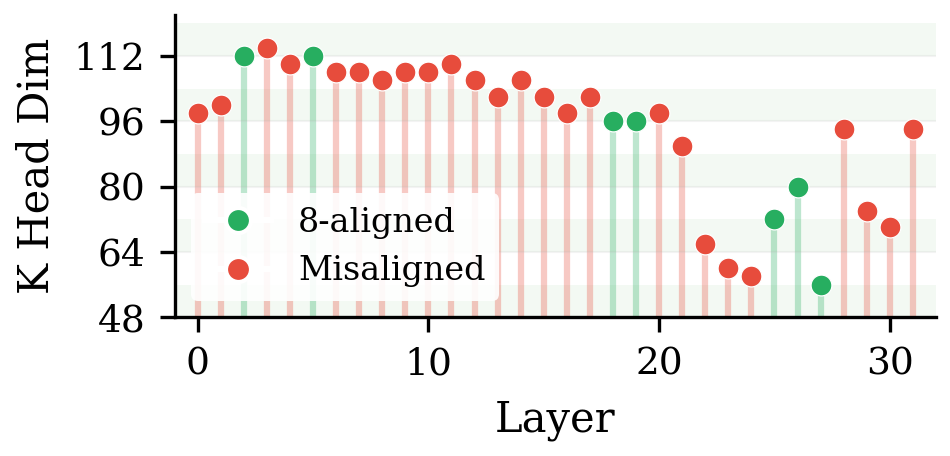}
\caption{Llama-3-8B after PaLU compression at $\rho=20\%$.}
\label{fig:palu_dist}
\end{figure}

We demonstrate the dimensional misalignment problem with a real-world example: PaLU~\citep{palu}.
We use \textbf{8-alignment} ($d \bmod 8 = 0$) as an example of alignment constraints.
When dimensions are not multiples of 8, latency can increase by up to 90\% (Figure~\ref{fig:sdpa_latency}); we detail this analysis in \S\ref{sec:analysis}.
Figure~\ref{fig:palu_dist} shows that a large fraction of layers end up misaligned (e.g., 78\% in this setup).
For generality, we summarize the mainstream importance score proxies into four categories (Table~\ref{tab:importance_scores}).
We empirically measure unconstrained dimension allocation on Llama-3-8B at $\rho{=}20\%$ with all four proxies; Figure~\ref{fig:dim_scatter} demonstrates that misalignment consistently occurs across methods. Appendix~\ref{app:scatter_ratios} shows the misalignment persists across compression ratios $\rho \in [10\%, 50\%]$.

\begin{table}[t]
\centering
\caption{Importance scores for budget allocation.}
\label{tab:importance_scores}
\small
\setlength{\tabcolsep}{2.5pt}
\begin{tabular}{@{}lll@{}}
\toprule
\textbf{Method} & \textbf{Score} & \textbf{Works} \\
\midrule
Magnitude & $\|W_i\|_F$ & SVD-LLM~\citep{svdllm2024} \\
Activation & $\|X_i\|_F$ & ASVD~\citep{asvd} \\
Gradient & $\left|\frac{\partial \mathcal{L}}{\partial h_i} \cdot h_i\right|$ & Taylor Pruning~\citep{taylor_pruning} \\
Fisher & $\mathrm{tr}(\mathbf{F}_i)$ & PaLU~\citep{palu} \\
\bottomrule
\end{tabular}
\end{table}


\section{Full-Stack Analysis}
\label{sec:analysis}
\begin{figure}[t]
  \vspace{0.2cm}
  \centering
  \begin{tikzpicture}[
      box/.style={draw=gray!60, rounded corners=5pt, minimum width=1.8cm, minimum height=1.2cm,
                  line width=0.8pt, font=\sffamily\footnotesize, align=center, inner sep=3pt},
      >=stealth, arrow/.style={->, line width=1.2pt, gray!55},
      lbl/.style={font=\sffamily\scriptsize, text=gray!40!black, above=1pt}
    ]
    \node[box, fill=cblue!18] (fw) at (0,0) {Framework\\[1pt]\scriptsize\color{gray!60!black}FlashAttention\\\scriptsize\color{gray!60!black}Math};
    \node[box, fill=corange!18] (lib) at (2.8,0) {Library\\[1pt]\scriptsize\color{gray!60!black}cuBLAS\\\scriptsize\color{gray!60!black}CUTLASS};
    \node[box, fill=cred!18] (hw) at (5.6,0) {Hardware\\[1pt]\scriptsize\color{gray!60!black}Tensor Core\\\scriptsize\color{gray!60!black}CUDA Core};
    \draw[arrow] (fw) -- (lib) node[midway, lbl] {Dispatch};
    \draw[arrow] (lib) -- (hw) node[midway, lbl] {Execute};
  \end{tikzpicture}
  \caption{PyTorch SDPA execution stack.}
  \label{fig:fullstack_overview}
  \end{figure}
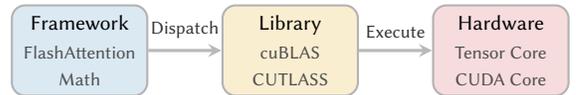

We analyze where and why dimensional misalignment causes slowdowns on an NVIDIA A100-80GB with PyTorch 2.9.1, CUDA 12.8, FP16.
Latency is measured with CUDA events (50 warmup, 200 measurement iterations, 3 trials).
Root causes fall into three layers (Figure~\ref{fig:fullstack_overview}): Framework, Library, and Hardware. We detail each layer in the following subsections.

\subsection{Framework Layer}
\label{sec:framework}
Existing DL frameworks such as PyTorch and TensorFlow dispatch an operation to different backends.
For example, a matrix multiply \texttt{A@B} in PyTorch may run via cuBLAS or via a Triton kernel depending on the shape and hardware, where the dispatching mechanism is hidden from the user.
In this section, we exemplify this issue via PyTorch's \textbf{SDPA} (Scaled Dot-Product Attention), the core attention mechanism:
\begin{equation}
\mathrm{Attention}(Q,K,V) = \mathrm{softmax}(Q K^\top / \sqrt{d_k}) V
\end{equation}
where $Q$, $K$, $V$ are the query, key, and value matrices and $d_k$ is the per-head dimension.
When one calls the SDPA API\footnote{\texttt{torch.nn.functional.scaled\_dot\_product\_attention(Q, K, V)}}, PyTorch may select an optimized implementation (e.g., FlashAttention) or fall back to a naive eager implementation (the ``Math'' backend).

\begin{figure}[t]
  \centering
  \includegraphics[width=\columnwidth]{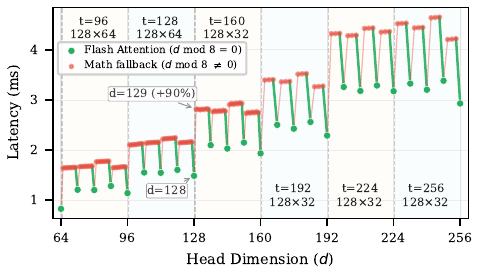}
  \caption{PyTorch SDPA latency across dimensions.}
  \label{fig:sdpa_latency}
  \end{figure}

We measured SDPA latency with inputs $Q,K,V$ of shape $(B, S, H, d)$: batch $B{=}4$, sequence length $S{=}2048$, number of heads $H{=}32$, and we sweep the per-head dimension $d$ from 64 to 256.
We observed a staircase pattern (Figure~\ref{fig:sdpa_latency}).
First, multiples of 8 are faster: e.g., $d{=}129$ incurs $\sim$90\% higher latency than $d{=}128$.
Profiling shows that PyTorch uses FlashAttention only when $d \bmod 8 = 0$; otherwise it falls back to the Math kernel.
Second, among 8-aligned dimensions, latency rises in a staircase at every 32-dimension boundary.
This is caused by FlashAttention-2 (FA2)'s template mechanism: FA2 selects the smallest template $t \geq d$ and assigns a tile shape $B_r {\times} B_c$ accordingly (Table~\ref{tab:fa2_templates}).
Crossing a template boundary (e.g., $d{=}128 \to 129$) halves $B_c$ and increases latency.
In Figure~\ref{fig:sdpa_latency}, alternating shades mark template regions; labels such as ``$t{=}96$, $B_r {\times} B_c{=}128{\times}64$'' show the template and tile shape.

\begin{table}[t]
\centering
\caption{FA2 template tiers and performance ($B{=}4$, $S{=}2048$, $H{=}32$).}
\label{tab:fa2_templates}
\small
\setlength{\tabcolsep}{3pt}
\begin{tabular}{@{}llrrr@{}}
\toprule
Region & Template & $B_r \times B_c$ & Latency & vs.\ $t{=}64$ \\
\midrule
$d{=}64$ & 64 & 128$\times$128 & 0.74\,ms & 1.0$\times$ \\
$d \in (64,96]$ & 96 & 128$\times$64 & 1.12\,ms & 1.5$\times$ \\
$d \in (96,128]$ & 128 & 128$\times$64 & 1.47\,ms & 2.0$\times$ \\
$d \in (128,160]$ & 160 & 128$\times$32 & 2.00\,ms & 2.7$\times$ \\
$d \in (160,256]$ & 192--256 & 128$\times$32 & 2.3--2.9\,ms & 3--4$\times$ \\
\bottomrule
\end{tabular}
\end{table}

\subsection{Library Layer}
\label{sec:library}

\begin{figure}[t]
\centering
\includegraphics[width=\columnwidth]{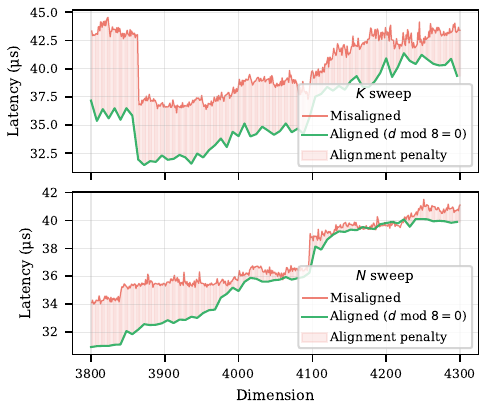}
\caption{GEMV latency ($M{=}1$, stride-1 sweep near 4096).}
\label{fig:gemv_alignment}
\end{figure}

\begin{figure*}[t]
\centering
\includegraphics[width=\textwidth]{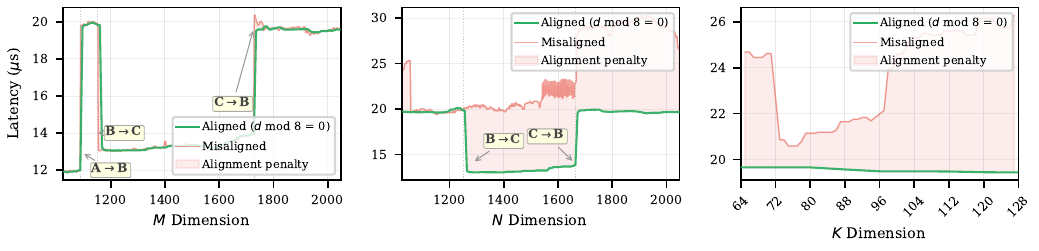}
\caption{GEMM latency with dimension sweep.}
\label{fig:GEMM_alignment}
\end{figure*}

Linear algebra is dispatched to libraries (e.g., cuBLAS), where the same GEMM can be served by different kernels depending on dimensions.
We examine this via GEMM $C {=} A {\cdot} B$ with $A \in \mathbb{R}^{M \times K}$ and $B \in \mathbb{R}^{K \times N}$.
We measured GEMM latency with two of $(M, N, K)$ fixed at typical LLM sizes ($M{=}N{=}2048$, $K{=}128$) and the third dimension swept from 50\% to 100\%.
Figure~\ref{fig:GEMM_alignment} shows the results.
First, $K$ and $N$ exhibit a clear alignment effect: when the swept dimension satisfies $d \bmod 8 = 0$, latency is lower (e.g., $K$ aligned $\sim$20\,$\mu$s vs.\ misaligned 22--26\,$\mu$s, up to 30\% penalty).
Second, $M$ and $N$ show \emph{kernel-switching cliffs}: at certain boundaries (e.g., $M{=}1728 \to 1729$), latency jumps (e.g., $\sim$30\%), where the cuBLAS heuristic selected an inefficient kernel.
We profiled with Nsight Compute to explain this: when $d \bmod 8 = 0$, cuBLAS invokes its native optimized kernel; otherwise it uses a CUTLASS-generated kernel, which is further divided into align2 (fetching 2 elements at a time) or align1.
Table~\ref{tab:cublas_tiers} summarizes the three tiers. 
GEMV ($M{=}1$) exhibits a similar but smaller penalty (${\sim}$12\% on $K$, ${\sim}$4\% on $N$; Figure~\ref{fig:gemv_alignment}), consistent with GEMV being memory-bound rather than compute-bound.

\begin{table}[t]
\centering
\caption{cuBLAS GEMM kernel tiers (Nsight Compute).}
\label{tab:cublas_tiers}
\small
\setlength{\tabcolsep}{2pt}
\resizebox{\columnwidth}{!}{%
\begin{tabular}{@{}clll@{}}
\toprule
\textbf{Tier} & \textbf{Condition} & \textbf{Kernel} & \textbf{MMA Instr.} \\
\midrule
\rowcolor{cgreen!10} 1 & $d \bmod 8 = 0$ & cuBLAS-native sm80 & \texttt{mma.m16n8k16} \\
\rowcolor{corange!10} 2 & $d \bmod 2 = 0$ & CUTLASS sm80 align2 & \texttt{mma.m16n8k16} \\
\rowcolor{cred!10} 3 & odd & CUTLASS sm75 align1 & \texttt{mma.m16n8k8} \\
\bottomrule
\end{tabular}%
}
\end{table}

\subsection{Hardware Layer}
\label{sec:hardware}
\begin{figure*}[t]
\centering
\includegraphics[width=\textwidth]{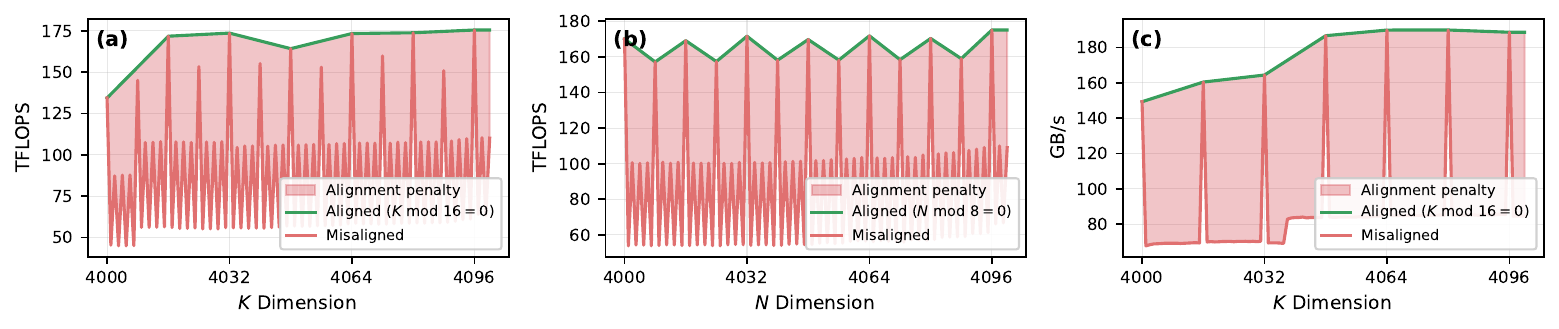}
\caption{Hardware-level alignment penalties (sweep near 4096): (a,b)~Tensor Core throughput, (c)~L2 Cache bandwidth.}
\label{fig:hw_alignment}
\end{figure*}

Beyond framework and library dispatch, misaligned dimensions also cause inefficiency from the hardware level.
We use Nsight Compute profiling to isolate two mechanisms.
\textbf{(1)~Tensor Core}: The A100 MMA instruction \texttt{mma.m16n8k16} processes tiles of $16{\times}8{\times}16$ fp16 elements; dimensions not divisible by these tile sizes leave partial tiles underutilized.
A throughput sweep near $K,N{=}4096$ confirms: aligned dimensions reach 160--175 TFLOPS while misaligned ones drop to 50--110 TFLOPS, with period-16 in $K$ and period-8 in $N$ matching the tile shape (Figure~\ref{fig:hw_alignment}a,b).
\textbf{(2)~Memory}: The A100 L2 Cache operates in 32-byte sectors; for FP16 this requires $K \bmod 16 = 0$ for full utilization.
Misaligned accesses show ${\sim}$2$\times$ bandwidth loss in microbenchmarks (Figure~\ref{fig:hw_alignment}c).

\paragraph{Summary.} Table~\ref{tab:constraints} summarizes all constraints.
The minimum requirement across all layers is $d \bmod 8 = 0$; stricter alignment (mod~16, mod~32) yields further gains.
These penalties compound: a single misaligned dimension can trigger a backend fallback, a suboptimal kernel, and underutilized tiles and memory accesses simultaneously.

\begin{table}[t]
\centering
\caption{Full-stack alignment constraints summary.}
\label{tab:constraints}
\small
\setlength{\tabcolsep}{2.5pt}
\begin{tabular}{@{}lllr@{}}
\toprule
\textbf{Level} & \textbf{Mechanism} & \textbf{Constraint} & \textbf{Penalty} \\
\midrule
\rowcolor{cblue!8} Framework & SDPA backend & $d$\%$8{=}0$ & ${\sim}$90\% \\
\rowcolor{cblue!8} Framework & FA2 template & $d$\%$32{=}0$ & ${\sim}$30\% \\
\midrule
\rowcolor{corange!8} Library & cuBLAS GEMM & $K$/$N$\%$8{=}0$ & ${\sim}$30\% \\
\rowcolor{corange!8} Library & cuBLAS GEMV & $K$\%$8{=}0$ & ${\sim}$12\% \\
\midrule
\rowcolor{cred!8} Hardware & TC MMA & $K$\%$16{=}0$, $N$\%$8{=}0$ & ${\sim}$70\% \\
\rowcolor{cred!8} Hardware & L2 sectors & $K$\%$16{=}0$ & ${\sim}$50\% \\
\bottomrule
\end{tabular}
\end{table}

\section{GAC: GPU-Aligned Compression}
\label{sec:gac}

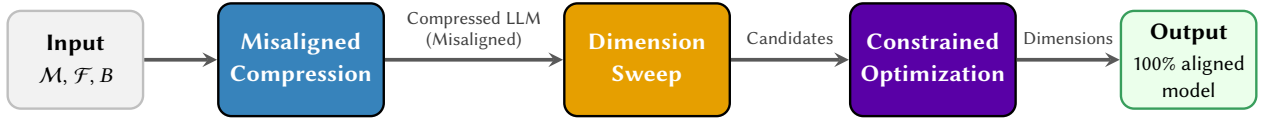
\begin{figure*}[t]
\centering
\begin{tikzpicture}[scale=1.0, every node/.style={scale=1.0},
    >=stealth,
    cblue/.style={fill={rgb,255:red,55;green,131;blue,187}},
    cred/.style={fill={rgb,255:red,211;green,63;blue,73}},
    cgreen/.style={fill={rgb,255:red,56;green,158;blue,92}},
    corange/.style={fill={rgb,255:red,230;green,159;blue,0}},
    phase/.style={draw, rounded corners=5pt, minimum width=2.2cm, minimum height=1.5cm,
                  line width=0.9pt, font=\sffamily\small, align=center},
    iobox/.style={draw, rounded corners=5pt, minimum width=1.8cm, minimum height=1.3cm,
                  line width=0.9pt, font=\sffamily\small, align=center},
    arrow/.style={->, line width=1.4pt, color=gray!70!black},
    lbl/.style={font=\sffamily\scriptsize, align=center, text=gray!50!black},
  ]

  \node[iobox, fill=gray!10, draw=gray!50] (input) at (0, 0) {
    \textbf{Input}\\[1pt]
    {\footnotesize $\mathcal{M}$, $\mathcal{F}$, $B$}
  };

  \node[phase, cblue, text=white] (compress) at (3.0, 0) {
    \textbf{Misaligned}\\[1pt]
    \textbf{Compression}
  };

  \node[phase, corange, text=white] (sweep) at (7.6, 0) {
    \textbf{Dimension}\\[1pt]
    \textbf{Sweep}
  };

  \node[phase, fill=violet!70!blue, text=white] (dp) at (11.4, 0) {
    \textbf{Constrained}\\[1pt]
    \textbf{Optimization}
  };

  \node[iobox, fill=green!8, draw={rgb,255:red,56;green,158;blue,92}] (output) at (14.8, 0) {
    \textbf{Output}\\[1pt]
    {\footnotesize 100\% aligned}\\[-1pt]
    {\footnotesize model}
  };

  \draw[arrow] (input) -- (compress);
  \draw[arrow] (compress) -- (sweep) node[midway, above=0pt, lbl] {Compressed LLM\\(Misaligned)};
  \draw[arrow] (sweep) -- (dp) node[midway, above=2pt, lbl] {Candidates};
  \draw[arrow] (dp) -- (output) node[midway, above=2pt, lbl] {Dimensions};

\end{tikzpicture}
\caption{GAC Pipeline.}
\label{fig:gac_framework}
\end{figure*}

\begin{algorithm}[t]
  \caption{GAC algorithm.}
  \label{alg:gac}
  \small
  \begin{algorithmic}[1]
  \Require Model $\mathcal{M}$, compressor $\mathcal{F}$, budget $B$, unit $u$
  \Ensure Aligned dimensions $\{d_i\}_{i=1}^n$ with $d_i \in C_i$
  \Statex \textit{Step 1: Unconstrained Compression}
  \State $\{d_i^*\}, \{s_i\} \gets \mathcal{F}(\mathcal{M}, B)$ \Comment{misaligned dims \& scores}
  \Statex \textit{Step 2: Dimension Sweep}
  \For{each weight $i = 1, \ldots, n$}
    \State Sweep aligned dims near $d_i^*$ using heuristic constraints from \S\ref{sec:analysis}
    \State $C_i \gets$ candidate aligned dims avoiding performance cliffs
  \EndFor
  \Statex \textit{Step 3: Constrained Optimization (Knapsack DP)}
  \State $B' \gets B / u$ \Comment{quantize budget}
  \State $D[0..n][0..B'] \gets -\infty$;\; $D[0][0] \gets 0$
  \For{$i = 1$ to $n$}
    \For{each $d_{ij} \in C_i$}
      \State $v_{ij} \gets s_i \cdot (|W_i(d_{ij})| {-} |W_i^*|)$;\; $w'_{ij} \gets |W_i(d_{ij})| / u$
      \For{$b = w'_{ij}$ to $B'$}
        \State $D[i][b] \gets \max\!\bigl(D[i][b],\; D[i{-}1][b{-}w'_{ij}] + v_{ij}\bigr)$
      \EndFor
    \EndFor
  \EndFor
  \State Backtrack from $\arg\max_b D[n][b]$ to get $\{d_i\}$
  \State \Return $\{d_i\}$
  \end{algorithmic}
  \end{algorithm}
  
To bridge the gap between compression and alignment, we propose \textbf{GAC} (GPU-Aligned Compression), a paradigm that makes budget allocation system-aware, so that fewer parameters translate into real speedup.
GAC wraps any dimension-reducing compressor with a post-processing step, re-selecting dimensions to satisfy alignment constraints.
Given a model $\mathcal{M}$ with $n$ compressible weights $\mathcal{W}$, a compressor $\mathcal{F}$, and a parameter budget $B {=} (1{-}\rho)\,|\mathcal{W}|$ ($\rho$: compression ratio), GAC produces a fully aligned model in three steps
(Figure~\ref{fig:gac_framework}).
Algorithm~\ref{alg:gac} details the GAC steps.

\subsection{Step 1: Misaligned Compression}
\label{sec:step1}

We first apply $\mathcal{F}$ to $\mathcal{M}$ without alignment constraints.
$\mathcal{F}$ can be any established dimension-reducing compressor---e.g., ASVD~\citep{asvd} (SVD factorization) or LLM-Pruner~\citep{llmpruner} (structured pruning).
Internally, $\mathcal{F}$ computes a per-weight importance score $s_i$ using one of the proxies in Table~\ref{tab:importance_scores} (e.g., activation magnitude for ASVD, gradient-based Taylor expansion for LLM-Pruner), then allocates dimensions $\{d_i^*\}$ proportionally: higher $s_i$ retains a larger $d_i^*$.
Because $s_i$ and the resulting $\{d_i^*\}$ are continuous, the compressed model is \emph{misaligned}.

\subsection{Step 2: Dimension Sweep}
\label{sec:step2}

A na\"ive fix would round every $d_i^*$ to the nearest sweet point (e.g. multiple of 8) from the Constraint Table~\ref{tab:constraints}.
However, we cannot hard-code alignment rounding rules because different operators exhibit different behavior across platforms (e.g., GPU architecture, PyTorch version).
A fixed heuristic that works on one platform may miss cliffs or exclude valid dimensions on another.

Instead, GAC selects candidates \emph{empirically}.
We use the heuristic constraints (e.g., $d \bmod 8 = 0$, $d \bmod 16 = 0$) to narrow the search space, then profile the kernel latency at each candidate near $d_i^*$ to verify it avoids performance cliffs on the \emph{actual} platform.
This produces a candidate set $C_i$.
For example, given $d_i^*{=}107.3$, the sweep yields $C_i{=}\{96, 104, 112, 128\}$: dimension 107 is excluded because it triggers a cuBLAS Tier-3 kernel (\S\ref{sec:library}), while 104 and 112 both land in Tier-1.
Because the sweep is hardware-specific, GAC adapts to different GPU architectures without manual tuning.

\subsection{Step 3: Constrained Optimization}
\label{sec:step3}

With misaligned dimensions $\{d_i^*\}$, importance scores $\{s_i\}$, and candidate sets $\{C_i\}$ in hand, we now select one aligned dimension per weight under the parameter budget.
Na\"ive rounding (e.g., round each $d_i^*$ to the nearest candidate) ignores two factors: (1)~different weights have different sensitivities, and (2)~rounding up at one weight consumes budget that could be spent elsewhere.
We therefore formulate a \emph{multi-choice knapsack} problem:
\begin{equation}
\max_{\{d_i\}} \sum_{i=1}^{n} s_i \cdot (|W_i| - |W_i^*|) \;\;\text{s.t.}\;\; \sum_{i} |W_i| \leq B,\;\; d_i \in C_i
\label{eq:gac}
\end{equation}
where $|W_i|$ is the parameter count of weight $W_i$ at dimension $d_i$, $|W_i^*|$ at the misaligned dimension $d_i^*$, and $C_i$ is the candidate set from Step~2.
The objective is \emph{asymmetric}: rounding up ($d_i > d_i^*$) preserves information (positive value), rounding down loses it (negative value), each scaled by the per-parameter importance $s_i$.
This lets high-importance weights receive more parameters while low-importance weights absorb the cost.

We solve Eq.~\ref{eq:gac} via dynamic programming.
For each candidate $d_{ij} \in C_i$, define value $v_{ij} = s_i \cdot (|W_i(d_{ij})| - |W_i^*|)$ and cost $w_{ij} = |W_i(d_{ij})|$.
The recurrence is:
\[
D[i][b] = \max_{j} \left\{ D[i{-}1][b - w_{ij}] + v_{ij} \right\}
\]
with complexity $O(n \cdot |C_{\max}| \cdot B')$, where $n$ is the number of weight matrices, $|C_{\max}| = \max_i |C_i|$ the largest candidate set, and $B'$ the quantized budget.

\noindent \textbf{Budget quantization.}
Na\"ively, the DP table has $B$ entries equal to the raw parameter budget, which can reach billions (e.g., $0.85 \times 100 \times 1024^2 \approx 10^8$ for 100 matrices of size $1024{\times}1024$ at $\rho{=}15\%$).
However, dimension reduction has a \emph{minimum cost unit}: pruning one column of a $[M,N]$ matrix by 1 adds or removes $M$ parameters.
If we further constrain dimensions to multiples of~8, the minimum unit becomes $u = 8 \cdot M_{\min}$, where $M_{\min}$ is the smallest row count across all weights.
We quantize costs and budget by $u$: $w'_{ij} = w_{ij} / u$, $B' = B / u$.
This reduces the DP table size dramatically---in the example above, $u{=}8{\times}1024{=}8192$ shrinks the table from ${\sim}10^8$ to ${\sim}12{,}500$ entries, a reduction of $8000{\times}$.
In practice, the DP runs in under one second on CPU---negligible compared to the compression itself.

\section{Evaluation}
\label{sec:eval}

\subsection{Setup}

We evaluate on Llama-3-8B with $\rho=15\%$, using PyTorch~2.9.1, CUDA~12.8, and FP16. Experiments were conducted on two architectures: NVIDIA Ampere (A100-80GB) and Hopper (H100-80GB). Unless stated otherwise, the results are collected from A100.
We also repeated a subset of the experiments (due to costs) on H100, which confirmed similar patterns (see Appendix~\ref{app:h100}).
We select two representative compressors that alter tensor dimensions in orthogonal ways:
\textbf{(1)~ASVD}~\citep{asvd}: activation-aware SVD ($W \to A \cdot B$) across all projection weights\footnote{All 32 layers $\times$ 7 projections (Q, K, V, O, gate, up, down) = 224 weights.}.
\textbf{(2)~LLM-Pruner}~\citep{llmpruner}: structured pruning of MLP weights\footnote{Layers 3--31 (29/32 layers); gate\_proj as pruning root, propagating to up\_proj and down\_proj.}.
We compare the uncompressed \emph{baseline} with \emph{Unaligned} (original compression) and \emph{GAC}.

\subsection{Implementation}

We implement GAC as a proof-of-concept by adding the DP solver on top of the existing ASVD and LLM-Pruner codebases (not yet a standalone framework; see \S\ref{sec:future}).
The dimension sweep profiles compression-sensitive kernels (e.g., GEMM, SDPA and GEMV) on the target GPU to build candidate sets (\S\ref{sec:step2}); the DP solver (\S\ref{sec:step3}) then selects aligned dimensions.
No model architecture changes, no runtime overhead, and no extra inference memory are required---GAC modifies only the dimension allocation before the final compression step.

\subsection{Preliminary Results}

Table~\ref{tab:main_results} summarizes the end-to-end comparison.

\begin{table}[t]
\centering
\caption{Preliminary results on Llama-3-8B ($\rho{=}15\%$). Measured with batch${=}1$, sequence length ${=}1024$.}
\label{tab:main_results}
\small
\setlength{\tabcolsep}{3pt}
\begin{tabular}{@{}lcrccc@{}}
\toprule
\textbf{Method} & \textbf{Align} & \textbf{PPL} & \textbf{PiQA} & \textbf{HSwag} & \textbf{Latency\,(ms)} \\
\midrule
\rowcolor{gray!10} Baseline & 100\% & 6.14 & 0.80 & 0.50 & 99.6 \\
\midrule
ASVD & 5\% & 34.7 & 0.58 & 0.28 & 100.5\,{\scriptsize\textcolor{cred}{(+1\%)}} \\
ASVD (GAC) & 100\% & 31.3 & 0.57 & 0.26 & 67.1\,{\scriptsize\textcolor{cgreen}{($-$33\%)}} \\
\midrule
LLM-Pruner & 83\% & 9.88 & 0.80 & 0.49 & 137.7\,{\scriptsize\textcolor{cred}{(+38\%)}} \\
LLM-Pruner (GAC) & 100\% & 9.87 & 0.78 & 0.47 & 88.0\,{\scriptsize\textcolor{cgreen}{($-$12\%)}} \\
\bottomrule
\end{tabular}
\end{table}

\noindent \textbf{Alignment.}
ASVD's unconstrained allocation produces 95\% misaligned dimensions, while LLM-Pruner produces 17\% misaligned dimensions (it only prunes MLP, so attention weights stay aligned).
GAC brings both to 100\%, snapping every dimension to an aligned candidate via the asymmetric DP objective (Eq.~\ref{eq:gac}).

\noindent \textbf{Accuracy.}
We report perplexity on WikiText-2 and accuracy on two tasks (PiQA, HSwag; 200 samples each).
For ASVD, GAC lowers PPL from 34.7 to 31.3 and the accuracy scores stay comparable
(PiQA 0.58$\to$0.57, HSwag 0.28$\to$0.26).
For LLM-Pruner, PPL is nearly identical (9.88 vs.\ 9.87) and downstream accuracy is well preserved (PiQA 0.80$\to$0.78, HSwag 0.49$\to$0.47), confirming that aligned re-selection does not sacrifice quality.

\begin{figure}[t]
\centering
\includegraphics[width=\columnwidth]{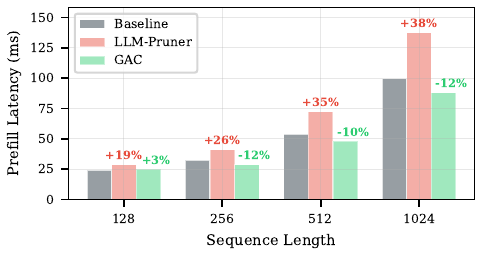}
\caption{Llama-3-8B latency across sequence lengths.}
\label{fig:prefill_scaling}
\end{figure}

\noindent \textbf{Latency.}
We measure prefill latency (batch${=}1$, $S{=}1024$; decode latency is left for future work).
Despite reducing parameters by 15\%, unaligned ASVD shows \emph{no speedup} (100.5\,ms vs.\ 99.6\,ms baseline)---the benefit is consumed by misalignment overhead.
GAC restores evident speedup (67.1\,ms, $-$33\%).
For LLM-Pruner, even 83\% alignment still incurs +38\% latency; GAC eliminates the penalty, achieving 12\% speedup over the \emph{uncompressed} baseline.
The penalty grows with sequence length (Figure~\ref{fig:prefill_scaling}): from +19\% at $S{=}128$ to +38\% at $S{=}1024$, as longer sequences push GEMMs deeper into the compute-bound regime where alignment (\S\ref{sec:hardware}) dominates.

\section{Related Work}
\label{sec:related}

\noindent \textbf{System-aware compression.}
Dimension-reducing compressors such as SVD factorization~\citep{asvd,svdllm2024,palu}, structured pruning~\citep{sparsegpt,wanda,llmpruner}, and KV eviction~\citep{h2o,quest,pyramidkv} optimize accuracy under a size budget but ignore how the resulting dimensions interact with GPU execution.
HALP~\citep{halp2021} and HALOC~\citep{haloc2023} incorporate hardware awareness into CNN compression, but treat latency as a \emph{black-box} signal: they optimize aggregate runtime without isolating \emph{why} certain dimensions are slow, and are tied to specific CNN architectures with no parameter-budget guarantee.
GAC instead identifies root causes at three levels (framework dispatch, kernel selection, hardware tile alignment; \S\ref{sec:analysis}) and constrains dimension selection directly, complementing any upstream compressor while guaranteeing both alignment and parameter budget.

\noindent \textbf{Serving-side mitigations.}
Serving systems handle misalignment \emph{reactively}.
FlashAttention-2 pads to the next template
(${\sim}$30\% overhead; \S\ref{sec:framework}); vLLM~\citep{vllm} rejects unsupported head dimensions.
These add overhead or break compatibility.
GAC prevents misalignment at compression time, eliminating such workarounds.

\section{Limitations and Future Work}
\label{sec:future}

\noindent \textbf{Framework automation.}
Our current implementation adds the GAC DP solver on top of the ASVD and LLM-Pruner codebases.
A fully general GAC framework---where users supply only a model name, compression ratio, and compressor, and receive a 100\%-aligned model---remains future work.

\noindent \textbf{Model coverage.}
We evaluate on a single dense model (Llama-3-8B).
Extending to more varieties (larger scales, MoE architectures, etc.) would test GAC's generality.

\noindent \textbf{Hardware diversity.}
Newer GPUs such as Blackwell impose stricter alignment~\citep{nvidia_tensor_core_evolution2024,nvidia_hopper_whitepaper,flashattention3};
Specialized devices, e.g., DGX Spark and Jetson Nano, add further constraints.

\noindent \textbf{Latency coverage.}
We report prefill latency at batch${=}1$, $S{=}1024$.
Covering a wider range of batch sizes and sequence lengths, as well as autoregressive decode latency, would give a fuller performance picture.

\noindent \textbf{Serving-engine compatibility.}
Our benchmarks use vanilla HuggingFace PyTorch models without any inference-time optimization.
Validating GAC under optimized serving engines (e.g., vLLM~\citep{vllm}, TensorRT~\citep{tensorrt}) and DL compilers (e.g., TVM) is needed to confirm alignment gains carry over to production stacks.

\clearpage
\bibliographystyle{ACM-Reference-Format}
\bibliography{references}

\clearpage
\appendix
\onecolumn
\section{Dimensional Misalignment Persists Across Compression Ratios}
\label{app:scatter_ratios}
\begin{figure*}[h]
\centering
\begin{subfigure}[t]{\textwidth}
\includegraphics[width=0.9\textwidth]{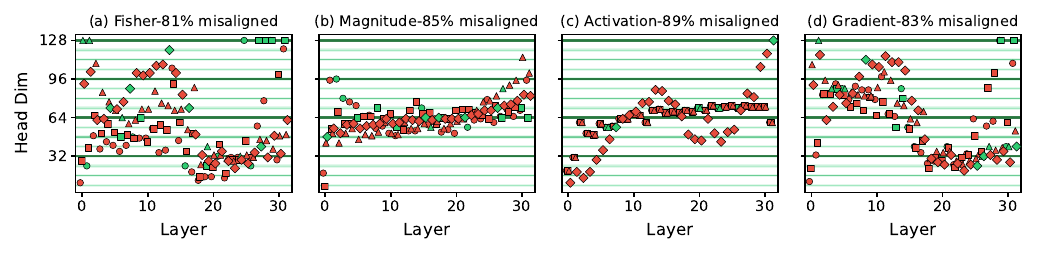}
\caption{Llama-3-8B, $\rho=50\%$}
\end{subfigure}
\vspace{-0.2cm}
\begin{subfigure}[t]{\textwidth}
\includegraphics[width=0.9\textwidth]{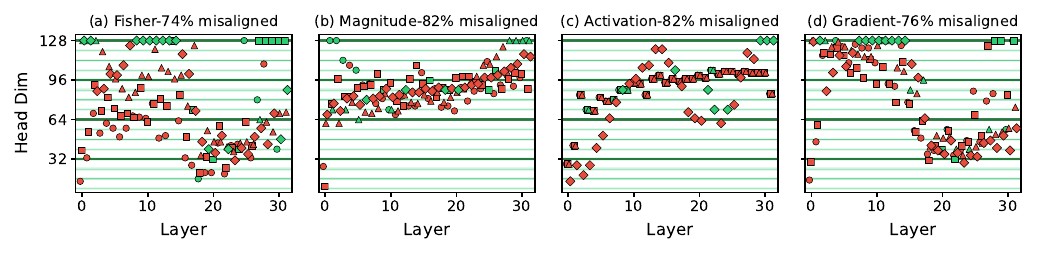}
\caption{Llama-3-8B, $\rho=40\%$}
\end{subfigure}
\vspace{-0.2cm}
\begin{subfigure}[t]{\textwidth}
\includegraphics[width=0.9\textwidth]{figures/scatter_1x4_meta_llama_3_8b_instruct_r0.8.pdf}
\caption{Llama-3-8B, $\rho=30\%$}
\end{subfigure}
\vspace{-0.2cm}
\begin{subfigure}[t]{\textwidth}
\includegraphics[width=0.9\textwidth]{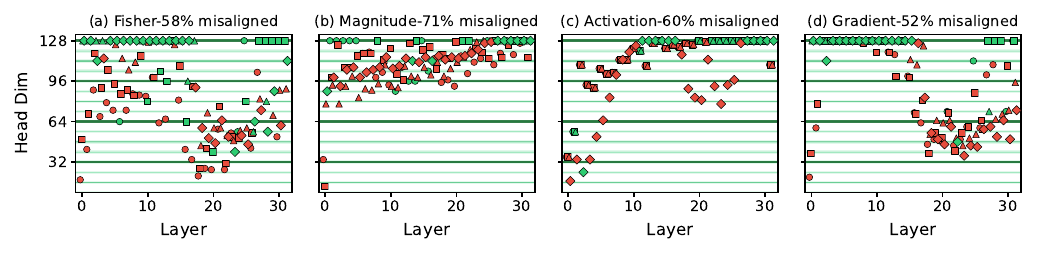}
\caption{Llama-3-8B, $\rho=10\%$}
\end{subfigure}

\caption{Compressed dimension distributions across compression ratios.}
\label{fig:scatter_ratios}
\end{figure*}

The misalignment problem persists across different compression ratios.
Figure~\ref{fig:scatter_ratios} shows dimension scatter plots for Llama-3-8B under unconstrained SVD allocation at four compression levels ($\rho{=}10\%$, $30\%$, $40\%$, $50\%$) using Fisher importance scores.
At every ratio, a substantial fraction of dimensions are misaligned, confirming that dimensional misalignment is inherent to importance-based rank allocation, not an artifact of aggressive compression.

\clearpage
\section{Dimensional Misalignment on H100 GPU}
\label{app:h100}
To demonstrate that dimensional misalignment is a fundamental hardware-software co-design issue and not specific to the Ampere architecture, we replicate our key profiling and evaluation experiments on an NVIDIA H100-80GB GPU (Hopper architecture). The Hopper architecture introduces new Tensor Core instructions (e.g., Tensor Memory Accelerator or TMA) and different cache hierarchies, which impose even stricter alignment constraints to achieve peak performance.

\begin{figure}[h]
\centering
\includegraphics[width=0.6\textwidth]{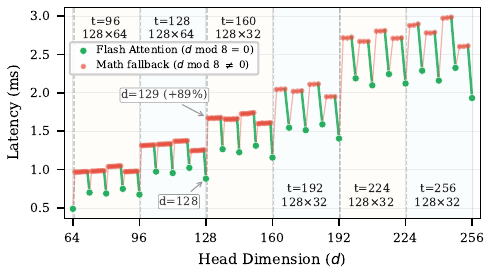}
\caption{PyTorch SDPA latency across dimensions on H100.}
\label{fig:sdpa_latency_h100}
\end{figure}

\textbf{Framework-Level Penalties.} Figure~\ref{fig:sdpa_latency_h100} shows the SDPA latency on H100. Similar to the A100 results, we observe a distinct staircase pattern. However, the penalties for misalignment are even more pronounced. Because the H100 has a significantly higher peak compute throughput, falling back to unoptimized kernels (e.g., the Math backend when $d \bmod 8 \neq 0$) results in a much larger relative slowdown compared to the A100.

\begin{figure*}[h]
\centering
\includegraphics[width=\textwidth]{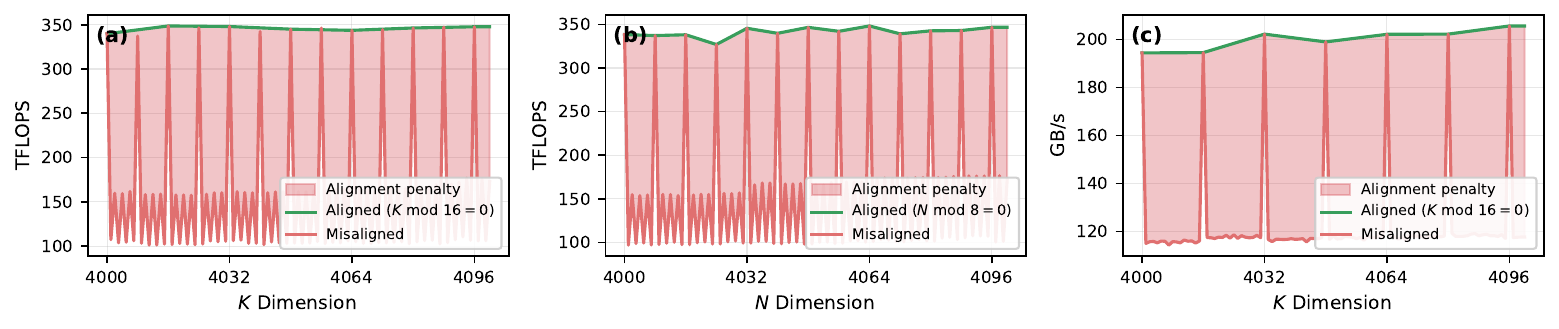}
\caption{Hardware-level alignment penalties on H100: (a,b) Tensor Core throughput, (c) L2 Cache bandwidth.}
\label{fig:hw_alignment_h100}
\end{figure*}

\textbf{Hardware-Level Penalties.} Figure~\ref{fig:hw_alignment_h100} details the hardware-level alignment penalties on H100. The impact on Tensor Core throughput (Figure~\ref{fig:hw_alignment_h100}a,b) and L2 Cache bandwidth (Figure~\ref{fig:hw_alignment_h100}c) follows the same periodic degradation patterns observed on Ampere, confirming that the underlying tile and sector alignment requirements remain critical bottlenecks for irregular dimensions.

\begin{figure}[h]
\centering
\includegraphics[width=0.6\textwidth]{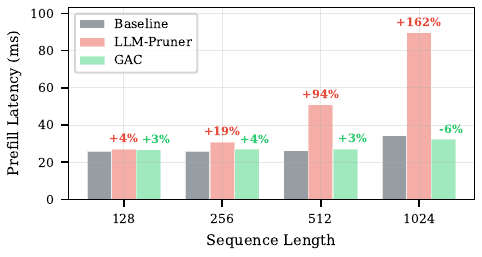}
\caption{Llama-3-8B latency across sequence lengths on H100.}
\label{fig:prefill_scaling_h100}
\end{figure}

\textbf{End-to-End Latency.} Finally, Figure~\ref{fig:prefill_scaling_h100} shows the end-to-end prefill latency scaling on H100 for Llama-3-8B. The unaligned compressed models suffer from severe slowdowns, completely negating the theoretical benefits of parameter reduction. In contrast, GAC successfully reduces latency by enforcing hardware-aligned dimensions, especially for larger batch sizes. This demonstrates that our alignment-aware compression paradigm is highly effective across different GPU generations. 

\end{document}